\documentclass{elsart}

\usepackage{subfigure}
\usepackage{graphicx}
\usepackage{alltt}
\usepackage{float}

\floatstyle{ruled}
\newfloat{code}{htbp}{loc}
\floatname{code}{Example}

\newcounter{bla}

\newcommand{\nin}{\noindent}

\newcommand{\bea}{\begin{eqnarray}}  
\newcommand{\eea}{\end{eqnarray}} 
\newcommand{\be}{\begin{equation}}  
\newcommand{\ee}{\end{equation}}

\begin{document}
\begin{frontmatter}

\title{LAW: A Tool for Improved Productivity with High-Performance Linear Algebra Codes.\\Design and Applications}

\author[a]{Timothy Stitt\thanksref{author}},
\author[b]{Graham Kells},
\author[b]{Jiri Vala}

\thanks[author]{email: timothy.stitt@ichec.ie}

\address[a]{Irish Centre for High-End Computing (ICHEC), NUI Galway, Ireland.}
\address[b]{Department of Mathematical Physics, NUI Maynooth, Ireland.}


\begin{abstract}
LAPACK and ScaLAPACK are arguably the defacto standard libraries among the scientific community for solving linear algebra problems on sequential, shared-memory and distributed-memory architectures. While ease of use was a major design goal for the ScaLAPACK project; with respect to its predecessor LAPACK; it is still a non-trivial exercise to develop a new code or modify an existing LAPACK code to exploit processor grids, distributed-array descriptors and the associated distributed-memory ScaLAPACK/PBLAS routines. In this paper, we introduce what we believe will be an invaluable development tool for the scientific code developer, which exploits ad-hoc polymorphism, derived-types, optional arguments, overloaded operators and conditional compilation in Fortran 95, to provide wrappers to a subset of common linear algebra kernels. These wrappers are introduced to facilitate the abstraction of low-level details which are irrelevant to the science being performed, such as target platform and execution model. By exploiting this high-level library interface, only a single code source is required with mapping onto a diverse range of execution models performed at link-time with no user modification. We conclude with a case study whereby we describe application of the LAW library in the implementation of the well-known Chebyshev Matrix Exponentiation algorithm for Hermitian matrices.

\begin{flushleft}

\end{flushleft}


\end{abstract}

\end{frontmatter}

\section{Introduction}

While the \textit{memory-wall effect}\cite{Wulf95}, which describes the growing disparity between CPU speed and memory latencies, has been well known in the high-performance computing (HPC) community for many years, only recently it seems has there been renewed interest\footnote{As exemplified by the DARPA High-Productivity Computing Systems Project - http://www.darpa.mil/ipto/programs/hpcs/} in the disparity between the ``\textit{time to solution}'' of production codes and the ``\textit{time to solution}'' of the respective code development. 


Throughout the history of supercomputing, the progress of high performance architectures has been relentless, while unfortunately their companion software tools and environments have generally failed to keep pace. Users of such systems, who are frequently HPC non-specialists, are typically presented with ever-evolving, non-intuitive, and complex parallel-programming tools, libraries and development environments, which can require valuable time and prior development experience, before they can be effectively exploited to attack the computational task at hand. 

To tackle the imminent crisis in HPC productivity versus HPC peak performance, it is imperative that high-level, user friendly, portable library tools are developed which abstract away the underlying system-dependent complexity and provide the user with a simple, yet powerful interface allowing them to concentrate solely on solving the scientific problem at hand, without being distracted by lower-level architectural issues. Furthermore, the scientific code developer should not be inconvenienced with maintaining separate code branches for each target machine. In the ideal case, a single code should be maintained which can be mapped onto serial, shared-memory or distributed-memory architectures at link time with no modification required to the original source.

While influential HPC libraries such as MPI, LAPACK 3.0, BLAS, ScaLAPACK and PBLAS\footnote{also commercial libraries such as NAG and Visual Numerics}, have succeeded in raising the abstraction bar, they can still arguably be considered as low-level tools, due in part to their mnemonic routine signatures, complicated invocation arguments and reliance on architecture-dependent features (for instance, during the initialization and invocation of message-passing frameworks), particularly by those users with limited development experience. While historically this development approach may have been unavoidable, modern programming languages possess advanced abstraction features such as modularity and polymorphism whereby current HPC library tool developers should be able to fully abstract away system dependent issues, leaving the developer to concentrate on the scientific domain details to which they are more comfortable and accustomed. For example, the PETSc library\cite{Petsc97} is an excellent example of a tool promoting abstraction which is developed with modern software engineering principles, whereby users can be experiment freely with different solvers (both serial and parallel) in the solution of systems modeled by partial differential equations. 

In this paper we introduce a set of high-level wrapper library routines to the high-performance BLAS and PBLAS libraries, which are independent of the target architecture, for a set of common linear algebra operations, by exploiting advanced features of Fortan 95 such as generic functions, derived-types, optional arguments, overloaded operators and source code preprocessing. By providing a single abstract user interface, developers are excused from considering a given architecture, whether it be serial, shared-memory multithreaded or distributed-memory parallel when invoking the routines and reasoning about their data storage semantics. It is believed such an environment will result in the rapid prototyping and development of scientific codes with automatic multi-architecture executables being generated during the linking phase, with minimal user interaction.

\section{Linear Algebra Libraries}\label{introduction}
LAPACK\cite{Lap99} and ScaLAPACK\cite{Scal97} and their respective lower level building block libraries, BLAS and BLACS/PBLAS, are arguably the defacto standard libraries among the scientific community for solving linear algebra problems on sequential and parallel architectures. While ease of use was a major design goal for the ScaLAPACK continuation project; with respect to its predecessor LAPACK; it is still a non-trivial exercise to develop a new code or modify an existing LAPACK code to exploit a processor grid, distributed-array descriptors and the associated distributed-memory ScaLAPACK routines. Furthermore, developing separate codes with individual LAPACK and ScaLAPACK routine instrumentation can lead to additional project management headaches including multiple code branches and code bloat. 

A novel approach to developing code with these high performance libraries would be to exploit a high-level library interface which disassociates the developer from the underlying architecture (whether it be serial or parallel) and allows them to concentrate on solving the scientific problem at hand without being concerned with cryptic routine names and their type associativity, numerous invocation parameters and dependence on a particular architecture (which may or may not require processor grids, blocking factors and distributed-array descriptors). To illustrate this goal, consider the routine invocations (written in Fortran syntax) given in Example 1, which perform a standard dense matrix-matrix multiply\footnote{of the form $C=\alpha AB + \beta C$} operation.\\

\begin{code}
\caption{Four Sample Dense Matrix-Multiply Routine Invocations}
\begin{alltt}
 \footnotesize{

1. CALL SGEMM(TRANSA,TRANSB,M,N,K,alpha,A(1,1),LDA,B(1,1),LDB,beta, &
                C(1,1),LDC)

2. CALL PSGEMM(TRANSA,TRANSB,M,N,K,alpha,A,1,1,DESCA,B,1,1,DESCB, &
                beta,C,1,1,DESCC)

3(a). CALL MATRIX_MULTIPLY(A,B,C,alpha,beta)

3(b). C = alpha*A*B + beta*C}
 \end{alltt}
\end{code}

Invocations $1$ and $2$ typify the calls required to perform a matrix-matrix multiply operation\footnote{For single-precision real-valued dense matrices which are not submatrices} within the sequential BLAS and parallel PBLAS libraries respectively. It should be remarked that the calling syntax of both these routines somewhat betray their functionality by confusing their identity with oversized arities and cryptic arguments. Furthermore, invocation 2 assumes a distributed matrix representation with appropriately initialized array descriptors while no such information is required in invocation 1. 

Invocation $3(a)$ illustrates an improved calling interface whereby no assumption is made about the system-dependent storage structure of the matrices (i.e. is the matrix distributed or non-distributed); the developer is only concerned with supplying the appropriate values that describe the mathematical procedure undertaken. While it might be argued that invocations 1 and 2 provide greater flexibility in describing the multiplication process (by allowing the selection of array slices and submatrices), it can easily be shown that optional arguments provide a more modern and flexible approach to customizing calling method $3(a)$ if required. For instance, the developer can override the default behaviour of the matrix-multiply procedure by augmenting the calling signature with optional arguments as illustrated in Example 2.\\

\begin{code}
\caption{Flexible Routine Invocation with Optional Arguments}
\begin{alltt}
 \footnotesize{

CALL MATRIX_MULTIPLY(A,B,C,alpha,beta,transposeA='yes', & 
    startRowB=5,startColumnB=10,rowsB=20,columnsB=30)}
 \end{alltt}
\end{code}

In this code segment the user has expressed their wish to multiply the transpose of matrix A with a 20 by 30 submatrix defined at row 5 and column 10 in the global matrix B. While these additional arguments unavoidably increment the arity of the calling signature, competing in size with the BLAS and PBLAS versions, it must be noted that optional arguments are not mandatory (by definition) and hence are not always required, particularly if the default behaviour of the matrix multiply process is satisfactory. While optional argument interfaces are not revolutionary in the scientific programming domain (as exhibited by existing libraries such as those provided by NAG, Visual Numerics and LAPACK v3.0), the LAW library aims to raise the abstraction bar even higher by ensuring that a statement of the form described by $3(a)$ is fully sufficient and complete to be invoked on any given system, irrespective of its execution model, with no further modification or instrumentation.

Furthermore, statement $3(b)$ in Example 1 illustrates an alternative, and arguably more natural calling interface for the matrix-multiply operation when the default behaviour of the multiplication is required i.e. no subarrays, slices or non unitary array steps are required in the definition of the matrix multiplication operation\footnote{in other words, multiplication is performed over the entire matrix array objects}. For invocation $3(b)$ to be realisable, the intrinsic operators $+$, $*$ and $=$ must to be defined over the data types describing the matrix objects $A$, $B$ and $C$. This \textit{overloading} of intrinsic operators is fully supported in Fortran 95 and has been exploited in the LAW library to provide the most natural interfaces for operations over matrices and vectors, again irrespective of machine architecture and execution model.

As motivation for the detailed description of the LAW library to follow, consider the program code given in Example 3.
\begin{code}
\caption{LAW Library Program Example}
\begin{alltt}
 \footnotesize{
program multiply

  use LAW

  implicit none

  type(real_2D_matrix)    :: A,B,C
  real,dimension(100,100) :: D
  logical                 :: successful

  ! Initialise LAW System
  call LAW_INITIALISE()

  call LAW_CREATE_MATRIX(A,100,100,successful,initialValue=1)
  call LAW_CREATE_MATRIX(B,100,100,successful,initialValue=2)
  call LAW_CREATE_MATRIX(C,100,100,successful)

  C = A * B
  D = C	

  ! Terminate LAW System and Program
  call LAW_EXIT(ok)

end program multiply
}
 \end{alltt}
\end{code}
This code leverages the LAW library to perform a trivial matrix-multiply operation\footnote{for dense matrices of rank 2 and size 100} across a diverse class of computing architectures. In this example LAW library routines\footnote{made accessible to the program via the \textit{use} clause} are invoked to create memory storage for three real-valued dense matrices A, B and C, each with shape 100 x 100. Optional arguments accepted by the \textit{LAW MATRIX CREATE} routine enable the matrices to be initialised to default values, in this case $1.0$ and $2.0$, for matrices $A$ and $B$ respectively. Finally a simple matrix-multiply operation is performed on matrices A and B with the resultant matrix stored in both the matrix $C$ and the the conventional array D.
 
While this task can easily be implemented using existing BLAS and PBLAS routines, the flexibility of the LAW library and its higher level abstraction interface becomes manifest when it can be shown that this single program source can exploit both serial and parallel distributed-memory systems (over $n$ processors) without modification. 

When compiled against the serial LAW library on a serial architecture, all matrices are implicitly represented using conventional serial array notation and serial BLAS xGEMM routines are invoked to perform the matrix multiplication. Alternatively, if the code is compiled against the parallel LAW library and executed on a distributed-memory architecture, then all matrices are implicitly partitioned and distributed block-cyclically\footnote{as required by ScaLAPACK's distribution scheme} across the available user-defined processor grid\footnote{grid configuration is read from a simple parameter file within the LAW Library's Initialisation routine} using user-supplied blocking factors. To perform the multiplication operation, a PBLAS PxGEMM routine is implicitly invoked with the resultant matrix stored in the distributed matrix $C$. Furthermore, in this example, the elements of the distributed matrix C are gathered and copied to the conventional 2D array $D$\footnote{ by exploiting the ScaLAPACK PxGEMR2D redistribution routine} on each node\footnote{the parallel code is executed in single program multiple data (SPMD) fashion}. Such an operation can be very useful for collecting together the elements of a distributed matrix as a single entity on a single node e.g. sequential I/O from a master node. 

Furthermore, the serial LAW library could easily exploit multiple threads of a symmetric multiprocessing (SMP) architecture by linking to vendor-supplied multithreaded BLAS routines, such as those provided by Intel's Math Kernel Library (MKL), during the linking phase of the application build. This process will be discussed further in \textsection{\ref{build}}.

In each case, overloaded operators and generic functions provide the power and flexibility to allow the LAW library routines to hide the lower-level details of matrix element storage, and the serial or parallel multiplication strategy. All that is required is for the library user to ensure that they link against the serial or parallel LAW library dependent on their required execution model. Overall, we believe that this abstraction approach is necessary, useful and achievable for current and future classes of HPC libraries. 

\begin{figure}
  \begin{center}
      \includegraphics[height=3in,keepaspectratio=true]{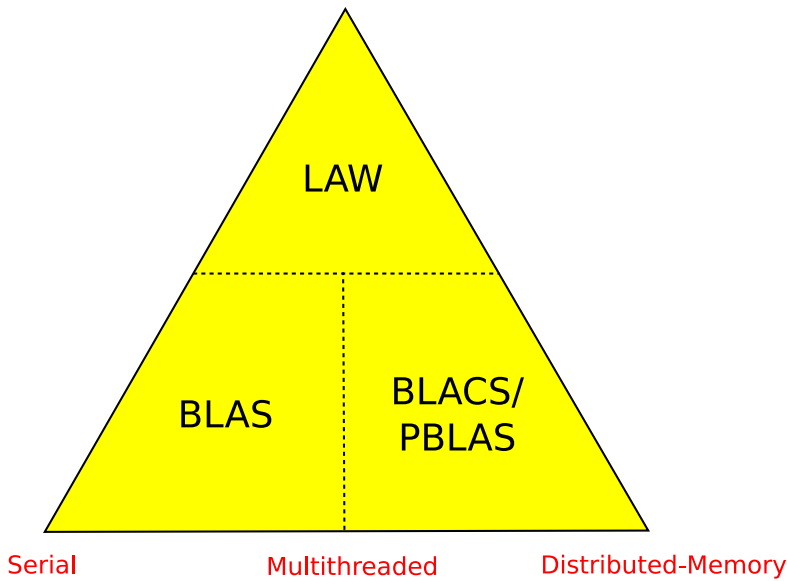}
    \caption{LAW Library Hierarchy}
    \label{Pyramid}
  \end{center}
\end{figure}

In summary Fig.~\ref{Pyramid} illustrates the hierarchical nature of the LAW library with respect to the lower level BLAS and PBLAS/BLACS libraries.
In the remainder of this paper we fully describe the design of the LAW library, and with the aid of a case study, illustrate its potential for the rapid prototyping and development of non-trivial linear-algebra based codes for execution on both serial and parallel architectures. It is hoped that this singular approach to interfacing with multi-platform high performance libraries will help promote abstraction within emergent scientific libraries. 

\section{The Linear Algebra Wrapper Library (LAW)}\label{LAW}
As discussed in \textsection{\ref{introduction}} the LAW library has been developed to provide a singular interface to a subset of common operations which can be found in the high performance BLAS and PBLAS linear algebra kernels. The motivating characteristics of this new interface is its independence from a given architectural model, whether it be sequential, multithreaded or distributed-memory parallel. To achieve this goal, it is imperative that abstraction be fully utilised in the design of the library's interfaces such that architecture-dependent features are successfully hidden  from the user. 

\subsection{Identifying abstraction in the BLAS and PBLAS libraries} 
Routines in the BLAS and PBLAS libraries can generally be classified by the following three attributes:
\begin{enumerate}
 \item \textit{Numerical Type} i.e. real, double precision, complex or double complex
\item \textit{Serial or parallel execution} (parallel PBLAS routine names are prefixed with the letter P) 
\item \textit{Functionality}
\end{enumerate}
Arguably, for most scientific users, only property (3) should be considered relevant in the development of a solution to a given task. In an ideal programming environment, the developer should not be concerned with data types and the underlying execution model. Dynamically-typed languages such as Python and Ruby are making significant progress in challenging traditional statically-typed languages such as C, C++ and Java in programming pedagogy\cite{Zelle99} as well as in the development of modern industrial codes due their  simpler type-devoid programming semantics. It is the authors view that HPC can benefit from these modern language developments, and when coupled with efficient numerical libraries will make a powerful and friendly programming environment for scientific code developers with limited programming experience (see \cite{Dong2006} for more details of recent research in exploiting the dynamically typed Python language in HPC). 

While it is impossible to protect the developer completely from matters of type in a statically-typed language such as Fortran, certain language features exist, like \textit{generic functions}, which can be exploited to provide a single unified interface, independent of argument types. For instance, in the matrix-multiply invocation $3(a)$ discussed in \textsection{\ref{introduction}}, the invoked routine name should be impartial to whether the matrices $A$, $B$ and $C$ are real-valued, complex-valued, double precision, dense or sparse. Ideally a single matrix-multiply routine should accept and manipulate input objects of varied yet sensible type. The LAW Library uses this concept of \textit{ad-hoc polymorphism} to simplify the library's programming interface. Furthermore, as will be shown in the following section,  distinction between sequential and parallel execution can be completely hidden by the LAW library by exploiting this self-same language feature. 

\subsection{The LAW Library Design Model} \label{LAWDesignModel}
\begin{figure}
  \begin{center}
      \includegraphics[height=3.5in,keepaspectratio=true]{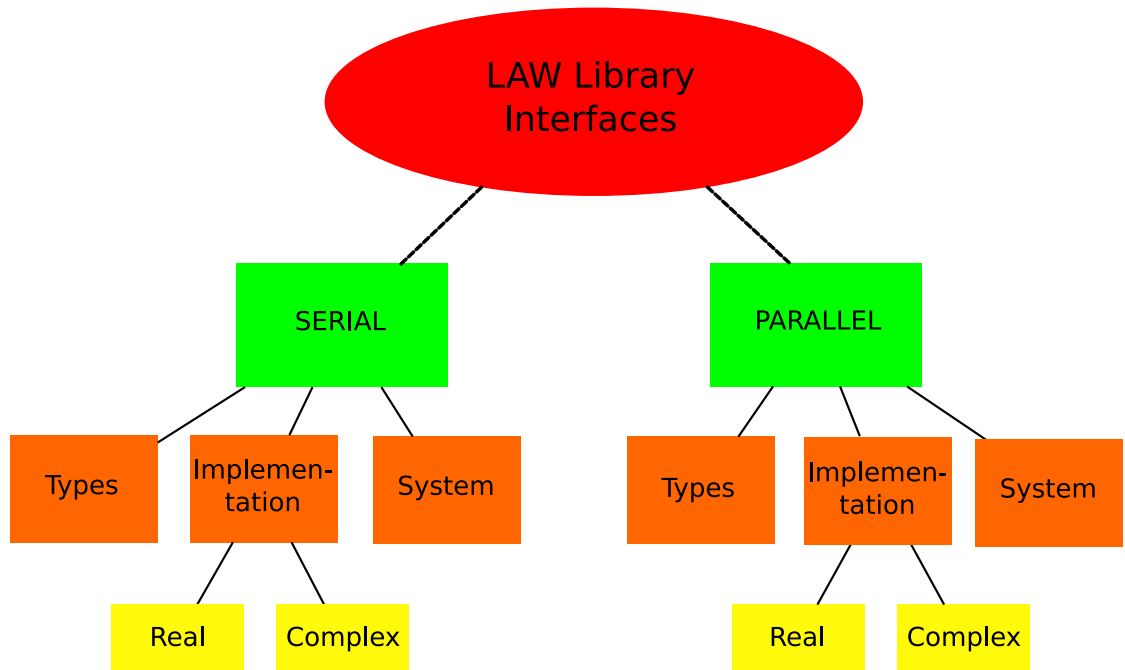}
    \caption{High Level LAW Library Design}
    \label{LAWDesign}
  \end{center}
\end{figure}

Fig.~\ref{LAWDesign} illustrates the top-level design characterising the LAW library. The library comprises a top-level interface which provides generic routine names for a common set of linear algebra operations involving matrices and vectors. 
This layer is at the highest level of abstraction and the routines at this level are independent of numeric type or execution model. Below this layer (and invisible to the user) are hidden the low-level concrete details of the library. At this sub-level the library differentiates between serial and parallel implementations and definitions of the lower level routines and types respectively. When a serial build of the library is requested, the high-level interfaces are mapped onto the concrete routines contained in the serial branch of the library. Likewise, during a parallel build the high-level LAW interfaces are mapped onto the implementations contained in the parallel branch of the library. 

Furthermore, the serial and parallel branches are composed of three (3) main components:
\begin{enumerate}
 	\item System Setup and Initialisation Routines
	\item Type Definitions
	\item Linear Algebra Routine Implementations
\end{enumerate}
Component (3) is further subdivided to contain real and complex implementations of the linear algebra routines.

\subsection{The System Routines}
The \textit{System} routines are required for initialising the LAW library's underlying framework for supporting either serial or parallel execution (and associated inter-node communication) i.e. whether it is hosting the BLAS or PBLAS versions of the linear algebra routines. The system sub-library consists of a set of constructor, accessor and mutator routines which are used to initialise, modify and grant access to architecture-dependent properties governed by the number of available processing elements, their ID's and respective topology. A collection of routines are also provided for collective data communication and barrier synchronisation among the available nodes. Fig.~\ref{SystemRoutines} summarises the generic routines available in the \textit{System} sub-library.

\begin{figure} 
\begin{center}
\begin{tabular}{||cc||}
\hline
LAW\_INITIALISE() & LAW\_EXIT() \\ 
LAW\_ID() & LAW\_NUMBER\_NODES()\\
LAW\_BROADCAST\_DATA() & LAW\_RECEIVE\_DATA()\\
\hline
\end{tabular}
\end{center}
\caption{Routines provided by the \textit{System} sub-library} 
\label{SystemRoutines} 
\end{figure}

When the LAW library is built for a serial architecture, minimal system setup and initialisation is required; effectively only a single node ID attribute is assigned indicating a solitary processing element. The total number of available nodes is also set to 1. 

When the LAW library is built for a distributed-memory parallel architecture a more complex initialisation is performed. In this environment, multiple processing elements may be available, associated under a particular topology\footnote{where all processors execute an identical copy of the code in a Single Program Multiple Data (SPMD) fashion}. In ScaLAPACK this topology is primarily grid-based. During initialisation, a grid context is created (using user supplied row and column grid parameters read from an external file\footnote{currently referred to as \textit{grid.config}}) and process ID's are assigned to all participating processors dependent on their location within the grid context. This initialisation process is implemented using the BLACS library which contains a rich set of routines for creating process grids and interrogating it's properties from within each processing element. 

To reinforce this implementation duality consider Examples 4 and 5 which illustrate the \textit{System} initialisation routines for both the serial and parallel branches of the LAW library.
\begin{code}
\caption{LAW\_INITIALISE() Routine for a Serial Architecture}
\begin{alltt}
 \footnotesize{

subroutine LAW_INITIALISE()
!------------------------------------------------------------------
! Description:
!   This routine initialises the framework for a serial 
!   architecture
!------------------------------------------------------------------
    node_ID=0             ! Set ID to 0 in a serial implementation
    total_nodes=1         ! Only one PE available
end subroutine LAW_INITIALISE}
 \end{alltt}
\end{code}
First and foremost, it should be noted that both routines maintain a consistent high-level interface yet their implementation details are different, as described in \textsection{\ref{LAWDesignModel}}. In effect, this allows the LAW library user to invoke the initialisation routine with the same routine name irrespective of target architecture. Only at link time is the correct implementation selected as will be described in \textsection{\ref{build}}.
\begin{code}
\caption{LAW\_INITIALISE() Routine for a Parallel Architecture}
\begin{alltt}
 \footnotesize{

subroutine LAW_INITIALISE()
    !-----------------------------------------------------------
    ! Description:
    !   This routine initialises the framework for a distributed 
    !   architecture
    !
    ! Local Scalars
    logical :: successful  
    !------------------------------------------------------------
    call initialise_grid(successful)
    if (.not.successful) then 
       print *,'Error Declaring Process Grid !'
       call LAW_EXIT(abort)
    end if
    
  contains

    subroutine initialise_grid(successful)
      !------------------------------------------------------------
      ! Description:
      !   This routine initialises a process grid and corresponding 
      !   context
      ! 
      ! Local Scalars  
      logical,intent(out) :: successful
      !------------------------------------------------------------
      call read_grid_data()    ! Read Grid Extents & Blocking Factors
      call blacs_get(-1,0,context)
      call blacs_gridinit(context,'Row',grid_rows,grid_columns)
      call blacs_gridinfo(context,grid_rows,grid_columns, &
         process_row,process_column)
      
      if (process_row==-1) then   ! Grid initialised unsuccessfully
         successful=.false.
      else
         successful=.true.
      end if
      call blacs_pinfo(node_id,total_nodes) ! Get Node ID & Size

    end subroutine initialise_grid

               .....}
 \end{alltt}
\end{code}

As a further example of the duality in the LAW Library System routines (for both serial and parallel branches) consider the \textit{LAW\_BARRIER()} routine given in Examples 6 and 7. Since synchronisation is redundant on a serial architecture with a single processing node, this routine performs no action\footnote{``smart'' compilers should be able to optimise 'away' this redundant invocation at compile time and therefore no performance penalty is accrued}. Conversely, the parallel barrier routine invokes the lower-level BLACS barrier routine and synchronises all available processing nodes at that point in the execution. 
 
\begin{code}
\caption{LAW\_BARRIER() Serial Implementation}
\begin{alltt}
 \footnotesize{

subroutine LAW_BARRIER()

end subroutine LAW_BARRIER}
 \end{alltt}
\end{code}

With these simple yet important routines the scientific code developer can develop a single program code which displays diverse behaviour when compiled and linked on different architectures. For example, consider the LAW program shown in Example 8. When linked against the serial LAW library, the master will perform some independent work and then a single salutation message will be displayed. In comparison, when linked against the parallel LAW library and executed on $p$ processors, $p$ salutation messages will be displayed, after the master process completes it's task.

\begin{code}
\caption{LAW\_BARRIER() Parallel Implementation}
\begin{alltt}
 \footnotesize{

subroutine LAW_BARRIER()

  call blacs_barrier(context,'A')

end subroutine LAW_BARRIER}
 \end{alltt}
\end{code}

\subsection{LAW Types} \label{LAWTypes}
The LAW library provides the user with a rich and powerful collection of aggregate types for declaring matrix and vector entities. Currently, one-dimensional (1D) vectors and two-dimensional (2D) matrix objects are supported across serial and parallel libraries. Recent updates have also seen the inclusion of a sparse matrix type\footnote{Using the Compact Storage Row representation (CSR)} for the serial library only and hence is not available for codes whose target platform is a distributed-memory architecture. Future updates to the LAW library hope to include complementary sparse matrix types and routines for distributed-memory machines\footnote{or wrappers to existing libraries such as PETSc that can provide parallel sparse matrix operations} to preserve the libraries portability across execution models.

\begin{code}
\caption{Sample LAW Code}
\begin{alltt}
 \footnotesize{
  program hello

      use LAW

      call LAW_INITIALISE()

      if (LAW_ID==0) then
          ....master does some work ....
      end if

      call LAW_BARRIER()

      print *,'Hello World'

      call LAW_EXIT(ok)

   end program hello}
 \end{alltt}
\end{code}

On serial architectures, the BLAS library represents matrix and vector objects as conventional 2D and 1D array declarations respectively (where array elements are stored contiguously in the memory-space using row-major distribution). On distributed-memory architectures, the memory associated with matrix and vector structures is partitioned and distributed across the available processing elements. To manage this data distribution, \textit{array descriptors} are associated with each distributed object, which records attributes describing the distribution of the global structure, including the global dimensions of the distributed object along with the row and column blocking factors. 

When declaring distributed entities, the ScaLAPACK developer is required to initialise the appropriate descriptors\footnote{ScaLAPACK provides routines to aid the developer in initialising these descriptors} and ensure that satisfactory memory storage is allocated and available on each processing node to contain it's locally distributed slice of the globally-distributed structure. As stated in \textsection{\ref{LAW}} the primary aim of the LAW library is to abstract away these architecture-dependent complexities and provide a uniform view of the execution and data storage model. In this context it is imperative that the user not be burdened with managing array descriptors if the target platform is a distributed-memory architecture. Furthermore, serial architectures require no reference to array descriptors and hence their declarations are redundant in a serial execution model. For these reasons, it is necessary that descriptors be hidden from the LAW library user at the library interface level. In principle the initialisation of array descriptors is entirely automatic assuming the user supplies sufficient information regarding the processor grid and row and column blocking factors\footnote{which are read from external file during parallel LAW initialisation stage}. This automatic generation of distributed arrays is tightly coupled to the creation of \textit{matrix} and \textit{vector} objects on distributed-memory architectures and hence can be implicit in the parallel LAW library creation routines, a process which will be fully described in \textsection{\ref{Implementations}}

Example 9 highlights the complex-valued matrix and vector derived types provided by the serial branch of the LAW library.
\begin{code}
\caption{LAW Serial Matrix and Vector Complex Aggregate Types}
\begin{alltt}
 \footnotesize{
type complex_vector
   complex(kind=precision),dimension(:),pointer   :: elements
   integer                                        :: length=0
end type complex_vector

type complex_2D_matrix
   complex(kind=precision),dimension(:,:),pointer :: elements
   integer                                        :: rows=0,columns=0
end type complex_2D_matrix

type complex_sparse_2D_matrix

   ! Represented in CSR Form

   integer,dimension(:),pointer                   :: rowPointers
   integer,dimension(:),pointer                   :: columnIndices
   complex(kind=precision),dimension(:),pointer   :: elements
   integer                                        :: rows=0,columns=0

end type complex_sparse_2D_matrix}
\end{alltt}
\end{code}
The vector and non-sparse matrix types contain pointers to vector and matrix elements which will be dynamically allocated during the execution of the LAW library's matrix and vector constructor routines discussed in \textsection{\ref{Implementations}}. Further attributes are provided in the derived types to record the extents of the objects shape\footnote{this allows easy referencing without appealing to Fortran's \textit{size} intrinsic function}. The sparse matrix type contains pointer attributes that reflect the compact storage row (CSR) nature of the sparse representation. Similarly, components are provided that record the global nature of the sparse representation (which are necessary for operations over the sparse matrix type). 

The \textit{precision} kind specifier can be set at compile-time to indicate either single or double precision. This kind specifier is referenced throughout the LAW library to provide consistent yet flexible multi-precision support within a single code.

Furthermore, the LAW library provides similar type definitions for real-valued matrices and vectors where the \textit{complex} Fortran data type has been replaced with the \textit{real} data type. Type names are also modified to indicate the \textit{real} nature of the type's attributes.

Example 10 shows the complex-valued matrix and vector aggregate types provided by the parallel branch of the LAW library.  
\begin{code}
\caption{LAW Parallel Matrix and Vector Complex Aggregate Types}
\begin{alltt}
 \footnotesize{
type complex_vector
   complex(kind=precision),dimension(:),pointer   :: elements
   integer,dimension(:),pointer                   :: descriptor
   integer                                        :: length=0
end type complex_vector

type complex_2D_matrix
   complex(kind=precision),dimension(:,:),pointer :: elements
   integer,dimension(:),pointer                   :: descriptor
   integer                                        :: rows=0,columns=0
end type complex_2D_matrix}
\end{alltt}
\end{code}
It should be observed that the only significant difference between serial and parallel matrix/vector types is the inclusion of the descriptor pointer,  which will reference the descriptor array associated with the distributed matrix/vector elements, as required by the PBLAS routines. Furthermore, it should be noted that the type names are identical to those provided by the serial LAW library. Again this reinforces the library's goal of abstraction by providing a uniform interface to the library's features irrespective of the implementation architecture. 

\subsection{LAW Interfaces} \label{Interfaces}
As described in \textsection{\ref{LAWTypes}}, the LAW library provides a rich selection of matrix and vector types with contrasting implementations on  serial and parallel architectures. No physical memory is associated with a type until an instance of the type has been declared or \textit{created}. For both matrix and vector types a set of creation routines may be required to accommodate every combination of type-value and representation. For instance a LAW matrix type can be either real or complex-valued and with either a dense or sparse representation. This requires a total of four  creation routine implementations, one for each possible combination of type and representation e.g.

\begin{enumerate}
 \item LAW\_CREATE\_2D\_REAL\_MATRIX()
 \item LAW\_CREATE\_2D\_COMPLEX\_MATRIX()
 \item LAW\_CREATE\_SPARSE\_2D\_REAL\_MATRIX()
 \item LAW\_CREATE\_SPARSE\_2D\_COMPLEX\_MATRIX()
\end{enumerate}

In this scenario, the LAW library user is provided with a flexible and powerful collection of creation routines, yet it can be argued such an approach can adversely affect code readability and overcomplicate the development experience by drowning the user in a sea of routine names. An alternative approach is to provide a single interface to a set of routines (with similar semantics) but which are distinguished by the arity and type of their input parameters. The number and types of the input arguments to the \textit{generic interface} determines the concrete implementation that is to be invoked. This mechanism is called \textit{ad-hoc polymorphism} and is facilitated in Fortran 95 by \textit{generic functions}. To illustrate the application of generic functions in the LAW library consider the code given in Example 11 and Example 12. 
\begin{code}
\caption{LAW Matrix Creation Interface}
\begin{alltt}
 \footnotesize{
interface LAW_CREATE_MATRIX
   module procedure LAW_CREATE_2D_REAL_MATRIX
   module procedure LAW_CREATE_2D_COMPLEX_MATRIX
   module procedure LAW_CREATE_SPARSE_2D_REAL_MATRIX
   module procedure LAW_CREATE_SPARSE_2D_COMPLEX_MATRIX
end interface}
\end{alltt}
\end{code}

In Example 11 a single LAW\_CREATE\_MATRIX() generic interface is defined and made available to the library user for the creation of matrix objects. 
An application of this interface is given in Example 12 whereby the user has requested two matrices with shape (100,100) to be created. 
\begin{code}
\caption{LAW Matrix Creation Sample Code}
\begin{alltt}
 \footnotesize{
type(real_2D_matrix)           :: A
type(complex_sparse_2D_matrix) :: B

call LAW_CREATE_MATRIX(A,100,100)
call LAW_CREATE_MATRIX(B,100,100)}
\end{alltt}
\end{code}

In both cases the LAW\_CREATE\_MATRIX() interface is invoked but at compile time a specific creation implementation is determined from the list (1)-(4) above based on matching the types of their input arguments to the arguments passed to the generic interface. For instance, the LAW\_CREATE\_2D\_\-REAL\_MATRIX() routine expects a matrix of type \textit{real\_2D\_matrix} to be passed in its first argument while the LAW\_CREATE\_2D\_COMPLEX\_\-MATRIX() routine expects a matrix of type \textit{complex\_sparse\_2D\_matrix} to be passed. In Example 12, the LAW\_CREATE\_2D\_\-REAL\_MATRIX() and LAW\_CREATE\_\-2D\_COMPLEX\_MATRIX() routines will be dispatched by the compiler to allocate matrices $A$ and $B$ respectively, via the LAW\_CREATE\_MATRIX() generic interface. Where possible, generic interfaces have been used to provide a single point of invocation for many of the linear algebra operations provided by the LAW library.

\subsection{Overloaded Operators}\label{overload}
In association with the generic interface facility is the ability to overload intrinsic operators and create new operators within the Fortran language. Overloading operators is achieved in a similar fashion to generic functions, whereby the unary or binary arguments to an operator are matched to a series of routines associated with the overloaded operator's interface. For instance, Example 13 illustrates the overloading of the addition (+) operator for matrix objects within the LAW library. 
\begin{code}
\caption{Overloaded LAW Addition Operator}
\begin{alltt}
 \footnotesize{
interface operator (+)
   module procedure LAW_COMPLEX_MATRIX_ADD
   module procedure LAW_REAL_MATRIX_ADD
end interface}
\end{alltt}
\end{code}
In this case both LAW\_COMPLEX\_MATRIX\_ADD() and LAW\_REAL\_MATRIX\_ADD() are subroutines expecting two matrix input arguments of the same type. Depending on the specified matrix type (whether real or complex), the compiler will dispatch the appropriate routine implementation. This is illustrated in Example 14, where the addition of two real matrices using the overloaded (+) intrinsic operator is illustrated.
\begin{code}
\caption{LAW Matrix Addition Sample Code}
\begin{alltt}
 \footnotesize{
type(real_2D_matrix)           :: A,B

call LAW_CREATE_MATRIX(A,100,100)
call LAW_CREATE_MATRIX(B,100,100)

A = A + B}
\end{alltt}
\end{code}
This overloading mechanism has been exploited liberally in the LAW library to provide more natural interfaces to operations over matrix and vector objects (and their combinations). As a further example the matrix-vector multiplication of a matrix $A$ and vector $X$ can be expressed in the following form:
\begin{verbatim}
        type (real_2D_matrix) :: A
        type (real_vector)    :: X,result

        result = A * x 
\end{verbatim}
In this example the overloaded (*) operator accepts a real-valued matrix and a real-valued vector, and performs a matrix-vector multiplication using an appropriate lower-level BLAS or PBLAS routine\footnote{typically xGEMV and PxGEMV}. Furthermore, the assignment operator (=) has also been overloaded within the LAW Library to accept a LAW vector (or matrix) and copy its attributes to another, previously declared vector (or matrix). 

\subsection{LAW Library Routine Implementations} \label{Implementations}

While the combination of generic functions and overloaded operators provides a flexible and powerful interface to the LAW library, the diversity of execution model is administered by the low-level LAW routine implementations. Consider the creation of matrix objects discussed in \textsection{\ref{Interfaces}}. Once a specific creation routine has been determined via the generic interface, contrasting actions are required depending on whether the mode of execution is serial or parallel. 

On a serial system conventional contiguous array storage is administered, while on a distributed-memory system, a block-cyclic distribution of the matrix is required, with the initialisation of an appropriate array descriptor. Consider the implementation of LAW\_CREATE\_2D\_REAL\_MATRIX() routine within the serial branch of the LAW library, as illustrated in Example 15.
\begin{code}
\caption{Real Matrix Creation Routine for Serial Architectures}
\begin{alltt}
 \footnotesize{
subroutine LAW_CREATE_2D_REAL_MATRIX(matrix,upper1,upper2,status, &
     initialValue)
  !----------------------------------------------------------------
  ! Description:
  !   This routine creates an instance of a 2D real-valued array
  !

  !use LAW_Types

  ! Subroutine Arguments:
  !  Scalars with Intent(in)
  integer,intent(in)                           :: upper1,upper2
  real(kind=precision),optional,intent(in)     :: initialValue
  !  Scalars with Intent(out)
  logical,intent(out)                          :: status
  !  Arrays with Intent(out)
  type(real_2D_matrix),intent(out)             :: matrix
  ! Local Scalars:
  integer                                      :: error_flag
  real(kind=precision)                         :: value
  !----------------------------------------------------------------

  status=.true.
  value=0_precision

  if (present(initialValue)) value=initialValue

  allocate (matrix%elements(1:upper1,1:upper2),stat=error_flag)
  if (error_flag==0) then
     matrix%elements=value           ! Initialise array elements
     matrix%rows=upper1
     matrix%columns=upper2
  else
     status=.false.
  end if

end subroutine LAW_CREATE_2D_REAL_MATRIX}
\end{alltt}
\end{code}
In this implementation, memory is dynamically allocated for the elements in the matrix as determined by the extents of the dimensions (labelled as \textit{upper1} and \textit{upper2}) passed as arguments. Furthermore, if the optional \textit{initialValue} argument is provided, then the matrix elements are initialised to the appropriate value. The extents of the matrix are recorded in the \textit{rows} and \textit{columns} components of the matrix object. Finally the modifications to the matrix object are made visible outside the routine via the \textit{intent(out)} specifier. 

Example 16 details the implementation of the LAW\_CREATE\_2D\_REAL\_\-MATRIX() routine for the parallel branch of the LAW library. 
\begin{code}
\caption{Real Matrix Creation Routine for Parallel Architectures}
\begin{alltt}
 \footnotesize{
subroutine LAW_CREATE_2D_REAL_MATRIX(matrix,upper1,upper2,status, &
    initialValue)
  !-----------------------------------------------------------------
  ! Description:
  !   This routine creates an instance of a distributed 2D 
  !   real-valued array
  
  ! Subroutine Arguments:
  !  Scalars with Intent(in)
  integer,intent(in)                          :: upper1,upper2
  real(kind=precision),optional,intent(in)    :: initialValue
  !  Scalars with Intent(out)
  logical,intent(out)                         :: status
  !  Arrays with Intent(out)
  type(real_2D_matrix),intent(out)            :: matrix
  ! Local Scalars:
  integer                             :: error_flag
  integer                             :: global_rows,global_columns
  integer                             :: local_rows,local_columns
  real(kind=precision)                :: value
  !-----------------------------------------------------------------

  status=.true.
  value=0.0_precision
  if (present(initialValue)) value=initialValue

  global_rows=upper1    ! Calculate Global Row/Column sizes
  global_columns=upper2

  ! Calculate Local Row/Col Sizes
  call LAW_LOCAL_MATRIX_SIZES(global_rows, & 
       global_columns,local_rows,local_columns)

  allocate(matrix%descriptor(9))        ! Create Descriptor Vector

  ! Initialise Descriptor 
  call LAW_INITIALISE_DESCRIPTOR(matrix%descriptor, & 
       global_rows,global_columns,local_rows)
  allocate (matrix%elements(local_rows,local_columns),stat=error_flag)

  if (error_flag==ok) then
     matrix%elements=value              ! Initialise array elements
     matrix%rows=upper1
     matrix%columns=upper2
  else
     status=.false.
  end if

end subroutine LAW_CREATE_2D_REAL_MATRIX}
\end{alltt}
\end{code}
In this implementation, a distributed array is constructed from the global matrix extents using appropriate routines provided in the LAW System sub-library. In this \textit{single-program multiple-data} (SPMD) parallel execution model each processor is assigned the responsibility of allocating sufficient local storage for its local assignments of the distributed array as determined by the blocking factors and defined processor topology. The low-level BLACS library provides a number of inquiry routines for determining the local row and column extents required by a given processing element, which is encapsulated in the LAW library's LAW\_LOCAL\_MATRIX\_SIZES() System routine. Once each processor determines its local matrix extents, local storage is allocated for the matrix elements. 

In addition, a distributed array descriptor of fixed size\footnote{currently ScaLAPACK requires a 1D array with 9 elements} is constructed for describing the attributes and structure of the distributed array\footnote{as required by PBLAS routines operating on distributed arrays}. Again, low-level PBLAS/BLACS routines are invoked with appropriate arguments to create the associated array descriptor via the LAW\_INITIALISE\_DESCRIPTOR() wrapper routine. Finally the matrix elements and attributes are initialised in the same fashion as the serial implementation.

To further highlight the diversification of serial and parallel implementations for a given routine within the LAW library consider the multiplication of two real matrices\footnote{expressed as C = A * B} implemented by the routine LAW\_REAL\_2D\_MATRIX\_\-MULTIPLY(). The implementation within the serial LAW library branch is given in Example 17.
\begin{code}
\caption{Real Matrix Multiplication for Serial Architectures}
\begin{alltt}
 \footnotesize{
function LAW_REAL_2D_MATRIX_MULTIPLY(A,B) result(C)
  !-------------------------------------------------------------------
  ! Description:
  !   This routine performs a matrix-matrix multiplication C = A * B 
  !   on serial architectures

  !use LAW_Types

  ! Subroutine Arguments:
  !  Arrays with Intent(in)
  type(real_2D_matrix),intent(in)            :: A
  !  Arrays with Intent(out)
  type(real_2D_matrix),intent(in)            :: B
  type(real_2D_matrix)                       :: C
  ! Local Scalars
  integer                                    :: m,n,k
  logical                                    :: successful
  !-------------------------------------------------------------------
  
  m = A%rows
  n = B%columns
  k = A%columns
  
  if (k /= B%rows) then
     print *,'Multiplied Matrices Do Not Conform!'
     call LAW_EXIT(abort)
  else

     call LAW_CREATE_MATRIX(C,m,n,successful);
     C = B
     C%rows=m
     C%columns=n
     
#if (PRECISIONFLAG==SINGLE)
     call sgemm('N', 'N', m, n, k, 1.0_precision, A%elements, m,  &
        B%elements, k, 0.0_precision, C%elements, m)
#else
     call dgemm('N', 'N', m, n, k, 1.0_precision, A%elements, m,  &
        B%elements, k, 0.0_precision, C%elements, m)
#endif
     
  end if

end function LAW_REAL_2D_MATRIX_MULTIPLY
}
\end{alltt}
\end{code} 

In this serial implementation, the extents of the contributing matrices are determined using the matrix's attributes contained in the derived type components. If the extents do not conform for multiplication then an error message is displayed and the code terminates in a controlled manner. If the extents conform then storage for the resultant matrix is created. Finally a low-level BLAS xGEMM routine is invoked to perform the matrix multiplication. The resultant matrix $C$ is then returned to the calling routine.

One point of note is the use of pre-processing directives within the matrix-multiplication routine for selecting between the scheduling of either \textit{single} or \textit{double} precision BLAS routines. During the build of the LAW library, the \textit{PRECISIONFLAG} macro variable is set by the user to indicate either single or double precision representation. This same mechanism is used to set the \textit{precision} kind specifier, which requests single or double precision for the real, complex and integer intrinsic types,  at the outset of the compilation\footnote{this process will be described in further detail in \textsection{\ref{build}}}.  If the PRECISIONFLAG is set to SINGLE, then all single precision BLAS (or PBLAS routines depending on whether a serial or parallel build is requested) will be selected, otherwise the alternative double precision routines will be invoked.
\begin{code}
\caption{Real Matrix Multiplication for Parallel Architectures}
\begin{alltt}
 \footnotesize{
function LAW_REAL_2D_MATRIX_MULTIPLY(A,B) result(C)
  !-----------------------------------------------------------------
  ! Description:
  !   This routine performs a matrix-matrix multiplication 
  !   on serial architectures

  ! Subroutine Arguments:
  !  Scalars with Intent(in)
  !  Arrays with Intent(in)
  type(real_2D_matrix),intent(in)            :: A
  !  Arrays with Intent(out)
  type(real_2D_matrix),intent(in)            :: B
  type(real_2D_matrix)                       :: C
  integer                                    :: m,n,k
  logical                                    :: successful

  external :: psgemm,pdgemm
  !------------------------------------------------------------------
  
  m = A%rows
  n = B%columns
  k = A%columns
  
  if (k /= B%rows) then
     if (LAW_ID()==0) then
        print *,'Multiplied Matrices Do Not Conform!'
     end if
     call LAW_EXIT(abort)
  else

     call LAW_CREATE_MATRIX(C,m,n,successful);
     C = B
     C%rows=m
     C%columns=n
     
#if (PRECISIONFLAG==SINGLE)
  call psgemm('N', 'N', m, n, k, 1.0_precision, A%elements,1,1, &
     A%descriptor, B%elements, 1,1, B%descriptor,0.0_precision, &
     C%elements, 1,1,C%descriptor)
#else
  call pdgemm('N', 'N', m, n, k, 1.0_precision, A%elements,1,1, &
     A%descriptor, B%elements, 1,1, B%descriptor,0.0_precision, &
     C%elements, 1,1,C%descriptor)
#endif
     
  end if

end function LAW_REAL_2D_MATRIX_MULTIPLY}
\end{alltt}
\end{code} 

In the parallel implementation of the real matrix-multiplication routine given in Example 18, the only distinction is that the BLAS xGEMM routine has been replaced by the equivalent PBLAS PxGEMM variant. Since all the attributes describing the matrix's structure (particularly the array descriptor) are packaged within the matrix derived type then minimal modifications need to be administered to generate the parallel PBLAS implementation from the serial BLAS template. All routines in the LAW library follow this fundamental structure, such that the parallel implementations can be obtained readily from the serial implementation. Most importantly, via generic interfaces and derived types, the underlying execution model is completely hidden, with no direct reference to array descriptors required by the library user. 

While it may be argued that this extra protective wrapper may generate additional invocation performance penalties, it is believed that such penalties are negligible and can be amortized during the execution particularly during intensive numerical computations where the linear algebra kernels dominate the wall-clock time. It is hoped that further development of the LAW library will help streamline the library's wrapper functions such that maximum performance can be achieved at little cost. We firmly believe though, that any performance penalties can be justified in light of the productivity benefits the LAW library can bestow to the scientific code developer and the subsequent rapid development of linear-algebra based codes across multiple architectures. 

Furthermore, the LAW library can be easily be extended by the library user to interface with additional BLAS/PBLAS routines that are currently not supported in the current version of the library. A standard structure is obeyed in the design of the library, as discussed in \textsection{\ref{LAWDesignModel}}, and new wrappers and associated implementations can easily be inserted with minimal effort. In addition, due to Fortran 95's lack of privatized derived type components\footnote{to be fixed in the new Fortran 200x release}, all matrix and vector type attributes can be directly accessed by the library user (when in scope) e.g. the user can directly reference matrix or vector elements by applying the derived type accessor operator $\%$. For instance, the elements of a vector can be explicitly intialised or interrogated as shown in Example 19.

\begin{code}
\caption{Extraction of Matrix Object Attributes}
\begin{alltt}
 \footnotesize{
    type(real_2D_matrix) :: A
    real(kind=precision) :: element

    A%elements(1) = 15.5
    element = A%elements(1)}
\end{alltt}
\end{code} 

Note that explicit element references on a parallel system refer directly to the local partition of the distributed matrix on the node making the reference. Hence accessing the first element in the locally distributed array, will most likely not refer to the first element in the global matrix/vector structure\footnote{apart from processor 0} and therefore this practice of explicit array reference should be approached with great care.

Notwithstanding this warning, experienced users can leverage the accessibility of the derived type components to interface to computational and driver routines in the full LAPACK and ScaLAPACK libraries. Since the problematic aspect of distributed array allocation is automatically performed during the creation of matrix and vector types, the library user can exploit the available type components in interfacing with additional LAPACK and ScaLAPACK routines above and beyond those available in the BLAS/PBLAS libraries. 

For example, a user can easily interface with the LAPACK and ScaLAPACK divide and conquer diagonalization routines \textit{xSYEDV} and \textit{xSYEDV}\footnote{which are not currently implemented in the LAW library} as shown in Example 20 (assuming prior creation of single precision matrices $A$ and $Z$ and appropriately allocated work arrays and parameters): 

\begin{code}
\caption{Interfacing with ScaLAPACK routines}
\begin{alltt}
 \footnotesize{
  call SSYEVD( JOBZ, UPLO, A%rows, A%elements, A%rows, W, WORK, &
   LWORK, IWORK,LIWORK, INFO )

  call PSSYEVD( JOBZ, UPLO, A%rows, A%elements, 1, 1, A%descriptor, &
   W, Z%elements, 1, 1, Z%descriptor, WORK, LWORK, IWORK, LIWORK, INFO)}
\end{alltt}
\end{code} 

\subsection{Array Redistribution}
On parallel systems array redistribution\footnote{invoked with the PBLAS redistribution routine PxGEMR2D} is a useful technique for converting a distributed array $A$, governed by blocking factors $r$ and $c$, into distributed array $B$, governed by blocking factors $r^{'}$ and $c^{'}$. A consequence of this process allows a conventional Fortran array $D$, with shape (\textit{rows},\textit{columns}) to be converted to/from a distributed array representation by associating $D$ with a row blocking factor of size \textit{rows} and column blocking factor of size \textit{columns} i.e. D of shape (rows,columns) is distributed by distributing blocks of size (row,column); ensuring that $D$ is distributed with all (row,columns) elements assigned locally to processor 0. This property allows a distributed array to be collected together on processor 0, or, a global array assigned locally to processor 0 to be distributed across the processor grid. These properties can be every useful when generating data at the master node with subsequent broadcasting to all other nodes, or for gathering data from all nodes back to the master for subsequent writing to file or master node processing. 

To illustrate these techniques consider Example 21.
\begin{code}
\caption{LAW Redistribution Example}
\begin{alltt}
 \footnotesize{
program processing

  use LAW

  implicit none

  type(real_2D_matrix) :: A
  real(kind=precision) :: data(1000,1000)
  logical              :: successful

  call LAW_INITIALISE(A,1000,1000,successful)
  
  ! Read data from file at master
  if (LAW_ID()==0) then
    open(123,file='old.data',form=unformatted)
    read(123) data
    close(123)
  end if

  call LAW_CREATE_MATRIX(A)
  
  A = D       ! Redistribute A on Master to all Processors

  .... some parallel processing of D ....

  D = A       ! Redistribute distributed array A back to D 
 
  ! Write data to file from master 
  if (LAW_ID()==0) then
    open(321,file='new.data',form=unformatted)
    write(321) data
  end if

  call LAW_DESTROY_MATRIX(A)
 
  call LAW_EXIT(ok)

end program processing
}
\end{alltt}
\end{code}     
In this program, the master node reads data from an external file to array $D$ on the master node. All processors then participate in the copy of D to the distributed matrix A via the overloaded assignment operator. Through the assignment interface, it can be determined that the copy is from the conventional array $D$ to the distributed matrix $A$. In this instance, the selected lower-level copy implementation routine constructs a \textit{dummy} array descriptor for $D$ treating it as a distributed matrix with a row blocking value of 1000 and a column blocking value of 1000 i.e. all elements are confined to master node (processor 0). 

The copy routine then performs a redistribution to the distributed matrix A (whose blocking factors are read from file during the LAW library initialisation phase at the outset of the program). At this point various linear algebra operations can be applied to A before the elements in the distributed matrix $A$ are gathered back to $D$ on the master node using a further redistribution from $A$ to $D$ (along with the dummy descriptor for $D$ as described in the previous assignment). The master node can now output the new elements of $D$ to external file.

When the program given in Example 21 is compiled against the serial LAW library no redistributions are required (since only the master node exists) and all matrix copies are undertaken via conventional array copies local to the master. 

\subsection{Data Value Broadcasting}
A relatively common task in parallel processing is for the master node to read some parameters from an external file and then broadcast those values to all other worker nodes participating in the calculation. The LAW library provides a wrapper to the BLACS broadcast routine xGEBS2D which can be exploited for this purpose. Consider the implementation of the LAW interface LAW\_BROADCAST\_DATA() given in Example 22.
\begin{code}
\caption{LAW Library Broadcast Routine}
\begin{alltt}
 \footnotesize{
subroutine LAW_REAL_DATA_BROADCAST(data)

  real(kind=precision),dimension(:),intent(in) :: data
  integer :: elements

  external :: SGEBS2D,DGEBS2D

  elements=size(data,dim=1)

#if (PRECISIONFLAG==SINGLE)
  CALL SGEBS2D(LAW_CONTEXT(), 'All', ' ', elements, 1, data, elements )
#else
  CALL DGEBS2D(LAW_CONTEXT(), 'All', ' ', elements, 1, data, elements )
#endif

end subroutine LAW_REAL_DATA_BROADCAST
}
\end{alltt}
\end{code}
 
In this routine it assumed that all data parameters to be broadcast are packaged contiguously as real-valued\footnote{a complementary complex-valued routine is also available for complex-based data} items in an array\footnote{similar to MPI message-passing}, which is passed to the routine. The data packet is then broadcast to all participating nodes via the xGEBS2D BLACS routine. Similarly when receiving a broadcast data packet, the BLACS collective receive routine xGEBR2D is invoked by all nodes to receive the packet as shown in Example 23. 

The data packet is then passed back to the calling routine for unpacking and processing. An example code segment illustrating the broadcast of parameters from the master node is given in Example 24.

In this example, the master reads two values from an external data file and packs them into the message buffer associated with the array $data$. If there is more than one node participating in the calculation i.e. the code is linked to the parallel LAW library, then the message is broadcast to all other nodes. If the execution model is serial then no broadcast is made and the master proceeds with the remaining calculation. In a parallel execution, the non-master nodes participate in a collective receive operation, where upon successful completion, all nodes have a copy of the message packet, which is then unpacked to its constituent parts. All nodes can then proceed with a copy of the data values generated at the master node. 

\begin{code}
\caption{LAW Library Receive Routine}
\begin{alltt}
 \footnotesize{
subroutine LAW_REAL_DATA_RECEIVE(data)

  real(kind=precision),dimension(:),intent(inout) :: data
  integer :: elements

  external :: SGEBR2D,DGEBR2D

  elements=size(data,dim=1)

#if (PRECISIONFLAG==SINGLE)
  CALL SGEBR2D(LAW_CONTEXT(), 'All', ' ', elements, 1, data, &
     elements, 0, 0 )
#else
  CALL DGEBR2D(LAW_CONTEXT(), 'All', ' ', elements, 1, data, &
     elements, 0, 0 )
#endif

end subroutine LAW_REAL_DATA_RECEIVE
}
\end{alltt}
\end{code} 

\begin{code}
\caption{LAW Library Broadcast/Receive Sample Code}
\begin{alltt}
 \footnotesize{
real(kind=precision) ::  data(2)

if (LAW_ID()==0) then

       ! Read Configuration Parameters when master node
       read(file,*) value1
       read(file,*) value2
       
       ! If more than one processor broadcast data to all nodes
       if (LAW_NUMBER_NODES()>1) then

          ! Pack data for broadcast
          data(1)=real(number_qubits)
          data(2)=real(numberTimeSteps1)
          
          ! Broadcast data message
          call LAW_BROADCAST_DATA(data)

       end if

    else

       ! All processors receive broadcast data from the master
       call LAW_RECEIVE_DATA(data)

       ! Unpack data values
       value1=int(data(1))
       value2=int(data(2))
       
    end if
}
\end{alltt}
\end{code}

\section{Building and Linking the LAW library}\label{build}
\subsection{Building}
It has been the goal of the previous sections to motivate the qualities of the LAW library as well as describe it design and implementation by exploiting generic functions, overloaded operators and conditional compilation. In particular, the pre-processing of macro variables within the LAW code is an invaluable mechanism in selecting between the scheduling of single or double precision BLAS/PBLAS routines. In addition, macro variables are also employed at the highest level of the LAW library, to select between serial and parallel branches of the library during the build phase.

As highlighted in Example 3, the LAW library is provided as a Fortran 95 module, which can be easily imported into a user's code via Fortran's \textit{use} clause. This controlling module uses pre-processing directives to determine the precision and serial/parallel execution model for the resultant library build. An summary snapshot of the controlling $LAW$ module is given in Example 25.  
\begin{code}
\caption{LAW Library Top-Level Driver Code}
\begin{alltt}
 \footnotesize{
module LAW

  ! Single or Double Precision
#if (PRECISIONFLAG==SINGLE)
  integer,parameter :: precision=kind(0.0)  ! Single Precision Kind
#else
  integer,parameter :: precision=kind(0.d0) ! Double Precision Kind
#endif

! Import Data Real and Complex LAW Data Types
#if (SYSTEM==SERIAL)
  include "./SERIAL/REAL/LAW_TYPES.f90"
  include "./SERIAL/COMPLEX/LAW_TYPES.f90"
#else
  include "./PARALLEL/REAL/LAW_TYPES.f90"
  include "./PARALLEL/COMPLEX/LAW_TYPES.f90"
#endif

! Import Real and Complex LAW Interfaces
#if (SYSTEM==SERIAL)
  include "./SERIAL/LAW_INTERFACES.f90"
#else
  include "./PARALLEL/LAW_INTERFACES.f90"
#endif

contains
  
#if (SYSTEM==SERIAL)
  include "./SERIAL/LAW_SYSTEM.f90"
  include "./SERIAL/REAL/VECTOR/LAW_VECTOR_ROUTINES.f90"
  include "./SERIAL/COMPLEX/VECTOR/LAW_VECTOR_ROUTINES.f90"
  include "./SERIAL/REAL/MATRIX/LAW_MATRIX_ROUTINES.f90"
  include "./SERIAL/COMPLEX/MATRIX/LAW_MATRIX_ROUTINES.f90"
  include "./SERIAL/REAL/MATRIX_VECTOR/LAW_MATRIX_VECTOR_ROUTINES.f90"
  include "./SERIAL/COMPLEX/MATRIX_VECTOR/LAW_MATRIX_VECTOR_ROUTINES.f90"
#else
  include "./PARALLEL/LAW_SYSTEM.f90"
  include "./PARALLEL/REAL/VECTOR/LAW_VECTOR_ROUTINES.f90"
  include "./PARALLEL/COMPLEX/VECTOR/LAW_VECTOR_ROUTINES.f90"
  include "./PARALLEL/REAL/MATRIX/LAW_MATRIX_ROUTINES.f90"
  include "./PARALLEL/COMPLEX/MATRIX/LAW_MATRIX_ROUTINES.f90"
  include "./PARALLEL/REAL/MATRIX_VECTOR/LAW_MATRIX_VECTOR_ROUTINES.f90"
  include "./PARALLEL/COMPLEX/MATRIX_VECTOR/LAW_MATRIX_VECTOR_ROUTINES.f90"
#endif
  
end module LAW
}
\end{alltt}
\end{code} 

It should be noted that assorted LAW library modules can be generated depending on the contents of the macro variables \textit{SYSTEM} and \textit{PRECISIONFLAG}. Currently both these variables are set in the LAW Library makefile\footnote{the user is provided with a GUI tool to register their choices}, and passed to the compiler using appropriate conditional compilation options. With these variables the compiler can build either the serial or parallel branches of the LAW library module by conditionally compiling the relevant sections (which encapsulate the appropriate selection of generic interfaces, types, system and low-level implementations), as dictated by the results of the conditional compilation selection tests. Furthermore, the precision kind specifier is also set via the \textit{PRECISIONFLAG} macro. In total, four combinations of execution model and type are allowable. It should also be noted that the low-level implementation routines for both serial and parallel branches are classified into either vector, matrix or matrix-vector routines and housed relative the top-level LAW module in individual source files. This placement format allows for easy identification and source-modification of the low-level routines when necessary. 

Finally, after a successful compilation (for choices of execution and precision model), the LAW library object code is packaged as a static library for eventual linking to a user's code. 

\subsection{LAW Library Usage}
Wrapper routines in the LAW library are imported into a user's code via Fortran 95's \textit{use} clause. During compilation of a user's code it is imperative that the include modules generated during the LAW library compilation are available to the compiler during compilation. This can be achieved by setting the appropriate include library path information in the user's code build instructions. Typically this is performed with the -I option on the compiler's command line e.g.
\begin{verbatim}
    $F90 -c foo.f90 -I /path_to_LAW_library/include
\end{verbatim}
Finally, both LAPACK and ScaLAPACK libraries should be linked against the appropriate LAW library during the linking phase. Sample linking commands are as follows:
\begin{verbatim}
    ! Serial Library Linking
    $(FC) -o foo foo.f90 -lserial_LAW -llapack -lblas

    ! Multithreaded Library Linking
    $(FC) -o foo foo.f90 -lserial_LAW -lmkl

    ! Distributed-Memory Library Linking
    $(FC) -o foo foo.f90 -lparallel_LAW -lblacs -lscalapack -lmpi
\end{verbatim}

 If a parallel build of the LAW library is linked then the user must supply a \textit{grid.config} parameter file to describe the the topology of the processor grid and the respective row and column blocking factors. Subsequent execution of the parallel code is then performed using a site-specific MPI execution command such as mpirun or mpiexec. If a multithreaded LAPACK library is linked against the LAW library then the user can experiment with the $OMP\_NUM\_THREADS$ environment variable prior to execution to exploit multiple threads during their calculation.  

\section{Case Study}

As an example of the the LAW library's versatility we study the well known Chebyshev matrix exponentiation routine for a Hermitian matrix. We aim to demonstrate how, with the aid of the LAW library, we can simplify the development of a single code source, that can be executed across a variety of different target architectures.

\begin{figure}
  \begin{center}
      \includegraphics[height=3.5in,keepaspectratio=true]{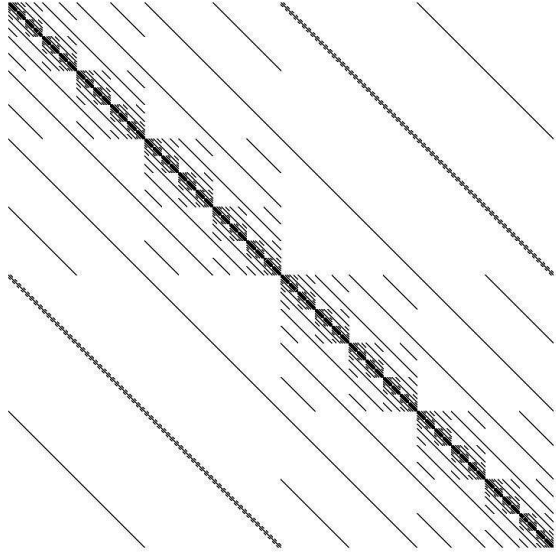}
    \caption{Structure of a 16-spin Hamiltonian matrix}
    \label{fig:H16V}
  \end{center}
\end{figure}

The matrix used, which we denote $\mathbf{H}$, is the Hamiltonian of a finite configuration of the Kitaev Honeycomb spin-1/2 lattice model \cite{kit05}. It is a real matrix but can be made complex by adding an external magnetic field along an appropriate direction. There are a number of different configurations which we may use with a varying number $(N)$ of spins. The matrices scale as $2^{N}$ x $2^N$ although they can also be given a sparse representation, see Fig.\ref{fig:H16V}. The model is important in the investigation into topological phases of matter and their application to topological quantum information processing.

The Chebyshev matrix exponentiation routine is well known \cite{kos94,ber99}. It can be used to generate the actual matrix exponential but is more commonly used to approximate the matrix-vector product using the Chebyshev polynomial expansion of a sparse Hermitian matrix. The main advantage of this routine is that one does not need to store the (usually) non-sparse matrix exponential. For this example we calculate $\exp(\tau \mathbf{H})\mathbf{v}$ for some real but negative $\tau$. The approximation takes the form 

\be
 \exp(\tau \mathbf{H})\mathbf{v}\approx \sum_{k=0}^m a_k C_k(\mathbf{B})\mathbf{v},
\label{eq:expand}
\ee

\nin where $C_k$ is the $k^{th}$ Chebyshev polynomial of the first kind and 

\be
\mathbf{B}=\frac{\mathbf{H}}{l_1}-\frac{l_2}{l_1}\mathbf{I}
\label{eq:Hscale}
\ee

\nin
with $\mathbf{I}$ being the identity matrix and

\be 
l_1=\frac{\lambda_n -\lambda_1}{2}, \quad l_2=\frac{\lambda_n+\lambda_1}{2}
\ee

\nin 
where $\lambda_n$ $(\lambda_1)$ are the maximum (minimum) eigenvalues of the original matrix $\mathbf{H}$. This scaling and shifting is required because the Chebyshev polynomials are only defined on the real axis between [-1,1] and the eigenvalues of any matrix to be expanded must be within this range. The coefficients $a_k$ are used to compensate and they are given by

\be
a_0=\exp( \tau l_2) I_0( \tau l_1), \quad a_k=2 \exp( \tau l_2) I_k( \tau l_1 )
\ee

\nin
where the $I_k$ are the modified Bessel functions. The integer $m$ in the sum in (\ref{eq:expand}) typically depends on $\tau$ and on the accuracy we require. Note that to compute the exponential $\exp(i \mathbf{H} \tau) \mathbf{v} $ a similar expansion can be used but using ordinary Bessel functions $J_k$. 

It should be noted that the estimate for $[\lambda_1,\lambda_n]$ should be from outside that range and that the convergence rates of the algorithm depends on its accuracy.  This problem of obtaining a good estimate for $[\lambda_1,\lambda_n]$ is addressed briefly in \cite{ber99}. 

The LAW Library code for performing the Chebyshev expansion phase of this algorithm is given in Example 26.
\begin{code}
\caption{Chebyshev Expansion LAW Library Code}
\begin{alltt}
 \footnotesize{
  type(complex_vector)    :: v1,v2,v3,v_initial,v_end
  type(complex_2D_matrix) :: B
  complex(kind=precision) :: alpha0,alpha1,alpha_k
  integer                 :: k,m
  logical                 :: successful

  call LAW_CREATE_VECTOR(v1,numberBasisStates,successful)
  call LAW_CREATE_VECTOR(v2,numberBasisStates,successful)
  call LAW_CREATE_VECTOR(v3,numberBasisStates,successful)
  call LAW_CREATE_VECTOR(v_initial,numberBasisStates,successful)
  call LAW_CREATE_VECTOR(v_end,numberBasisStates,successful)
  call LAW_CREATE_2D_MATRIX(B,numberBasisStates,numberBasisStates,successful)

  ! Normalise Hamiltonian
  B = (1/l1)*H + (-l2/l1)*I

  v1    = v_initial
  v2    = B * v1
  v3    = v1
  v_end = v2 
  
  alpha0 = LAW_GET_VECTOR_ELEMENT(alphas,1)/2
  alpha1 = LAW_GET_VECTOR_ELEMENT(alphas,2) 

  v_end = alpha0*v3 + v_end*alpha1

  !-- Main for loop for implementation of Chebyshev exp(-tau*H)

  do k = 2,m

    v3 = B * v2
    v1 = 2.0*v3 + (-1.0)*v1

    alpha_k = LAW_GET_VECTOR_ELEMENT(alphas,k+1)

    v_end = alpha_k*v1 + v_end

    v3 = v1
    v1 = v2
    v2 = v3

  end do 

  v_initial = v_end
}
 \end{alltt}
\end{code}

The mathematical objects fundamental to the propagation algorithm are the complex vectors \textit{$v1$, $v2$, $v3$, v\_initial, v\_end} and the complex normalized Hamiltonian $B$. Since the size of these objects are a function of the basis size, it is advantageous that we apply a solution that can exploit multiple processing elements and the large memory of a distributed-memory system if it is available. For this reason, we model these objects as flexible LAW Library matrix and vector types.

As discussed in \textsection{\ref{Implementations}} we create instances of the vector and matrix objects using the generic LAW Library creation routines \textit{LAW\_CREATE\_VECTOR()} and  \textit{LAW\_CREATE\_MATRIX()}. For vectors the creation routine requires the number of elements in the vector as an argument while the matrix creation routine requires the number of global rows and global columns in the matrix. The ancillary error flag labelled $successful$ can be interrogated after each creation routine to determine if the dynamic memory allocation was successful or not.

As described in \textsection{\ref{LAW}} the flexibility of the LAW library routines becomes apparent when you link against either the serial or parallel branches of the library. If the program source is linked against the parallel LAW library then all created vectors and matrices are distributed across a processor grid (as determined by the grid topology and blocking factor parameters set in the \textit{grid.config} file). When the source is compiled and linked against the serial LAW Library then no data-distribution is employed and a conventional array allocation is performed in the single memory-space.

Prior to propagation, we obtain the normalised Hamiltonian given in Eq.(2) above\footnote{For the sake of clarity we will assume $H$, $I$, $l1$ and $l2$ have been previously declared and initialised}. This normalisation expression introduces matrix addition via the overloaded LAW operators (*) and (+). In the expression
\begin{verbatim}
    B = (1/l1)*H + (-l2/l1)*I
\end{verbatim}
the matrices $H$ and $I$ are first scaled (since multiplication has a higher precedence than addition, in operator application). In a parallel system the scaling is performed across all processing elements and their associated local matrix partitions using a lower level parallel PBLAS scaling routine for performance. On a serial system the scaling is performed using a conventional BLAS scaling routine. The resultant scaled matrices then become input to the overload addition (+) operator where corresponding elements in each matrix are summed using lower level PBLAS or BLAS vector addition routines. The derived type attributes of the resultant matrix are then copied across to the left-hand side matrix $B$.

 The algorithm proceeds with a series of vector assignments. The overloaded assignment operator (=) provided by the LAW library will ensure that the derived type attributes of the objects on the right of the assignment will be copied to the corresponding attributes on the left of the assignment, as discussed in \textsection{\ref{overload}}. Special note should be made of the assignment to vector $v2$. Before the assignment completes, a matrix-vector multiply operation is performed via the overloaded product operator (*). Once again, the concrete matrix-vector multiply implementation is dependent on the branch of the LAW library that is linked. If the serial library is incorporated, then the matrix-vector multiply is performed serially using the $xGEMV$ BLAS routine. For the parallel LAW library, the product is performed in a distributed fashion across the processor grid using the corresponding parallel PBLAS $PxGEMV$ routine. 

Once the assignments are complete, the $alpha$ coefficients are computed as described in Eq.(4)\footnote{Within the code, the vector $alphas$ contains pre-computed coefficients where $alphas(k) = 2 \exp( \tau l_2) I_k( \tau l_1 )$}. The retrieval of a single value from a vector object can be accomplished with the \textit{LAW\_GET\_VECTOR\_ELEMENT()} library function. By supplying the global index of the required vector element as an argument to the routine, that element is retrieved irrespective of whether the element is stored locally or on a remote processing element (if the parallel LAW library is used). The calculation of the exact location of the global index is performed implicitly by the routine and the returned value is broadcast to all calling nodes.\footnote{In practice, the $alphas$ vector is small enough to not require distribution across the nodes in a parallel system, and hence does not need to be represented as a LAW vector object. We took the liberty to use a vector object in order to introduce the \textit{LAW\_GET\_VECTOR\_ELEMENT()} routine} 

The remainder of the propagation phase is carried out using combinations of the LAW library routines and functions already described. It should be clear from this example how a single version of the Chebyshev Matrix Exponentiation algorithm can be constructed, such that will execute on both serial and parallel systems without any modification to the source, by exploiting the high-level generic interfaces provided by the LAW library. As discussed in \textsection{\ref{build}}, multithreaded BLAS routines can be used in conjunction with the serial LAW library to incorporate the performance of multiple threads which are available on Symmetric Multiprocessing (SMP) systems.

\section{Conclusions}

In this paper we have introduced the LAW Library and showed that by exploiting advanced programming constructs in Fortran 90+ we can implement a set of high-level generic interfaces that allow the development of portable codes based upon BLAS and PBLAS linear algebra kernels. We believe this approach provides a number of benefits:
\begin{itemize}
 \item Rapid-prototyping of scientific codes which exploit BLAS/PBLAS calls
 \item Automatic handling of processor grids and array descriptors for distributed-memory systems
 \item Automatic redistribution of distributed matrices an overloaded assignment operator
 \item Only a single code source is required for diverse target architectures
 \item The library is easily extensible to more linear algebra routines and operations
 \item Simpler and more natural interface to standard linear algebra kernels
\end{itemize}
With the aid of a case study we have shown how LAW Library routines can simplify the development of non-trivial linear algebra calculations with a more natural and readable interface. Future developments will ensure that the set of matrix and vector operations is expanded for both serial and parallel systems under a single generic calling interface. This will include the provision for both serial and parallel sparse vector and matrix operations.

We hope that the LAW library will be found to be an invaluable tool for parallel code developers, in particular those with little or no previous experience with parallel communications and message-passing frame-works.
 
\section*{Acknowledgments}

The authors wish to acknowledge the SFI/HEA Irish Centre for High-End Computing (ICHEC) for the provision of computational facilities and support. J.V. and G.K. also acknowledge the Science Foundation Ireland (SFI) for support under the President of Ireland Young Researcher Award 05/YI2/I680.


\begin{thebibliography}{99}

\bibitem{Wulf95} Wulf, Wm. A. and McKee, \textit{Sally A. Hitting the memory wall: implications of the obvious.} ACM SIGARCH Computer Architecture News.  Volume 23 ,  Issue 1  (March 1995)

\bibitem{Lap99}  Anderson, E. and Bai, Z. and Bischof, C. and
                Blackford, S. and Demmel, J. and Dongarra, J. and
                Du Croz, J. and Greenbaum, A. and Hammarling, S. and
                McKenney, A. and Sorensen, D. \textit{LAPACK Users' Guide.} $3^{rd}$ Edition. Society for Industrial and Applied Mathematics. 1999. Philadelphia, PA. 0-89871-447-8 

\bibitem{Zelle99} Zelle, J. M. \textit{Python as a First Language.} Proceedings of the 13th Annual Midwest Computer Conference, March 1999.

\bibitem{Scal97} Blackford, L. S. and Choi, J. and Cleary, A. and
                D'Azevedo, E. and Demmel, J. and Dhillon, I. and
                Dongarra, J. and Hammarling, S. and Henry, G. and
                Petitet, A. and Stanley, K. and Walker, D. and
                Whaley, R. C. \textit{ScaLAPACK Users' Guide.} Society for Industrial and Applied Mathematics, 1997, Philadelphia, PAA, 0-89871-397-8 

\bibitem{Dong2006} Luszczek, P. and Dongarra, J.  \textit{High Performance Development for High End Computing with Python Language Wrapper (PLW)}. The International Journal of High Performance Computing Applications, Volume 21, No. 2, Summer 2007

\bibitem{Petsc97} Satish Balay and William D. Gropp and Lois Curfman McInnes and Barry F. Smith. \textit{Efficient Management of Parallelism in Object Oriented Numerical Software Libraries}. Modern Software Tools in Scientific Computing. Birkh{\"{a}}user Press. 1997

\bibitem{kit05} Kitaev A. {\em Anyons in an exactly solvable model and beyond} Ann.Phys. 321: 2-11 (2006)

\bibitem{kos94} Kosloff R. {\em Propagation methods for quantum molecular dynamics} Annu. Rev. Phys. Chem.  45:145-78 1994

\bibitem{ber99} Bergamaschi L. and Vianello M. {\em Efficient computation of the exponential operator for large, sparse, symmetric matrices, } Numer. Linear algebra Appl.{\bf } 7:27-45 (2000)

\end{thebibliography}
\end{document}